\documentclass[12pt,preprint]{aastex}

\begin{document}

\title{\bf H1517+656: the Birth of a BL~Lac Object?}

\author{Matthew O'Dowd}
\affil{School of Physics, University of Melbourne, Parkville, Victoria 3010, Australia; {\em modowd@physics.unimelb.edu.au}}
\author{C. Megan Urry}
\affil{Department of Physics and Yale Center for Astronomy and Astrophysics, P.O. Box 208121, New Haven, CT 06520-8121, USA; {\em meg.urry@yale.edu}}
\author{Riccardo Scarpa}
\affil{European Southern Observatory, Alonso de Cordova 3107, Vitacura, Casilla 19001, Santiago, Chile; {\em rscarpa@eso.org}}
\author{Randall B. Wayth}
\affil{School of Physics, University of Melbourne, Parkville, Victoria 3010, Australia; {\em rwayth@physics.unimelb.edu.au}}
\author{Rachel L. Webster}
\affil{School of Physics, University of Melbourne, Parkville, Victoria 3010, Australia; {\em rwebster@physics.unimelb.edu.au}}

\begin{abstract}

H1517+656 is an unusual source, even for a BL~Lac object. It is one of
the most luminous BL~Lacs known, with extreme emission properties
at radio, optical, and X-ray frequencies. Furthermore, in our WFPC2
snapshot survey we discovered a series of peculiar arcs describing a
2$^{\prime\prime}\!\!$.4 radius ring surrounding the source. This paper
describes follow-up observations with additional WFPC2 bands and the
STIS longpass filter, which have revealed this structure to be the
remnants of a very recent galaxy merger. Population synthesis
modelling has shown that regions of the arcs have stellar populations
with age $<$20~Myrs.  Additionally, the circularity of the arcs
indicates that the plane of the collision and hence accretion is very 
close to the plane of the sky.  Given that BL~Lac jets are thought to
be aligned with the line of sight, this observation may provide a direct link
between the transfer of angular momentum in an interaction
and the generation of a radio source. 

\end{abstract}


\section{Introduction}

The physical processes behind the BL~Lacertae phenomenon are not fully
understood. The presence of a relativistic jet closely aligned with
the line of sight is well accepted, but it is unclear whether this is 
sufficient to explain the near-absence of emission or absorption lines
in their spectra (see Urry \& Padovani 1995 for a review). It may be
that this lack of line features, as well as the intrinsically low
accretion rates of BL~Lacs compared to more powerful radio sources
\citep{ODowd}, 
is also due to a relatively low supply of gas to the broad line
region and accretion disk. Such a shortage may result from the
conditions of formation and environment of BL~Lacs, or perhaps be a 
function of their age as active radio sources. 

The {\it HEAO} 1 A-3 BL~Lac object H1517+656 is an extreme example of its
class, and may present a unique chance to probe some of these questions.
Not only is it the most optically luminous BL~Lac known, but
this extreme emission extends to both radio and X-ray frequencies. 
Although a high-frequency peaked BL~Lac, with an X-ray flux of
$f_X=2.89\times 10^{-11}$~erg cm$^2$ s$^{-1}$ (ROSAT 0.07 -- 2.4
keV band; Brinkmann \& Siebert 1994), H1517+656 was first detected in
the NRAO Green Bank 4.85 GHz radio catalog, with a flux density of $39
\pm 6$~mJy \citep{Becker}. 
Its optical brightness was found to be similarly high, first measured at
$B=15.5$~mag by \citet{Elvis}, where it was also identified as a 
BL~Lac. A lower limit on the redshift of H1517+656 of $z \ge 0.702$
was determined by \citet{Beckmann}, based on FeII and MgII
absorption lines.
Even at this minimum redshift, H1517+656 is currently the most optically
luminous BL~Lac object known, with 
$M_R\lesssim -27.1$~mag and $M_B\lesssim -26.3$~mag (assuming
$H_0=70$~km~s$^{-1}$~Mpc$^{-1}$, $\Omega_M=0.3$, $\Omega_{\lambda}=0.7$). 
Its X-ray and radio luminosities are also among the highest known for BL~Lacs,
with $L_X = 6.4 \times 10^{46}$~erg~s$^{-1}$ in the ROSAT PSPC band 
and $L_R = 0.82 \times 10^{26}$~W~Hz$^{-1}$ at 1.4 GHz from VLA
measurements of its unresolved core \citep{Kollgaard}.

\begin{figure}
\figurenum{1}
\plotone{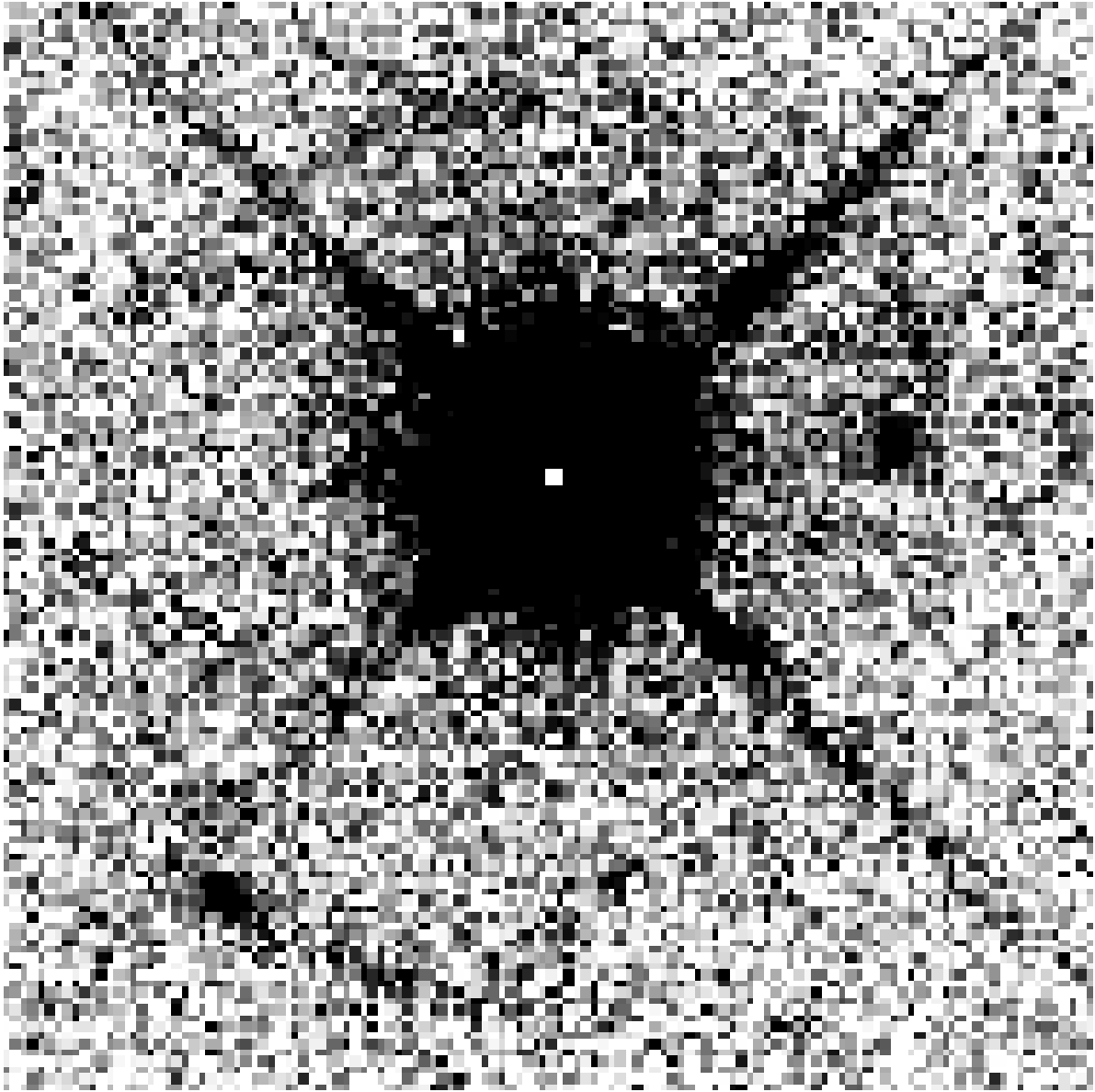}
\caption{
Cycle 6 snapshot image of H1517+656 in the F702W filter (approx. R band). 
\label{f702w1517}
}
\end{figure}

Due to the optically bright nucleus, efforts to detect the
host galaxy of this BL~Lac have been 
unsuccessful. Although not revealing the host galaxy, the
short HST snapshot R-band image from our Cycle 6 study \citep{Urry, Scarpa1}
uncovered an intriguing new mystery surrounding H1517+656:
a system of arcs and knots 
in a near-circular ring of radius 2$^{\prime\prime}\!\!$.4 around the
bright nucleus (Fig.~\ref{f702w1517}).

The average surface brightnesses of the arcs are 
$\mu_{R} \sim 22.4$~mag/arcsec$^2$, with radially-resolved 
widths of $\sim 0^{\prime\prime}\!\!.2$.
There are two especially bright 'knots' and position angles
126$^{\circ}$ and 260$^{\circ}$, which have magnitudes 
$m_{R} = 23.6$~mag and 23.8~mag, respectively.
The short, 320-second WFPC2 F702W exposure achieved a signal-to-noise
ratio of 1.5 per pixel in the arcs. While this provides a confident
detection, much additional detail is likely still to be undetected.
As a result, the origin of this structure could not be determined with
this data.  

As discussed in \citet{Scarpa1}, a number of different explanations for
the nature of the arcs present themselves. These fall into three main 
groups: either a) the arcs are gravitationally lensed images of an
extended background source; b) the arcs are physically associated
with the BL~Lac object, either as nearby structure or as part of the
host galaxy; or c) the arcs are the spiral arms of an intervening
face-on spiral galaxy. This last explanation is perhaps the least
likely, as discussed in \citet{Scarpa1}, however both (a) and (b)
warrant careful consideration.

In the case of (a), the only possible configuration is an extended
background source, either a galaxy or cluster of galaxies, being
lensed by either BL~Lac's host galaxy or local
group of galaxies. Were the BL~Lac part of the lensed object then, as
the central image, it would be heavily de-magnified, and a second image
would be observed outside the ring. This scenario had been investigated 
by \citet{Beckmann} based on the original HST data, and they find a
lensing mass of $>2\times10^{12}$M$_{\odot}$ is needed to produce arcs
of the observed radius.

In the case of (b), it is probable that the arcs are the remnants of 
a galaxy---galaxy interaction rather than structure within the BL~Lac
host galaxy. The diameter of the ring is at least $30$~kpc (at
the minimum redshift of the BL~Lac, $z=0.702$) --- too large to be the
spiral arms of a host galaxy \citep{Scarpa1}. If this scenario is 
correct, then we may be looking at a violent ongoing galaxy
interaction. If such interactions are indeed the primary triggers of 
activity in galaxy nuclei, then H1517+656 may be a BL~Lac in the midst
of its formation.
This is an interesting possibility given the extreme emission
properties of the source.

In this paper we describe the follow-up HST imaging program undertaken 
to determine the nature of the arcs surrounding H1517+656. In
section~\ref{obs} we describe the observations and data reduction,
including the host galaxy analysis 
and photometry of the arcs. We present our analysis and discussion in 
\S~\ref{1517disc}, where we describe why the interacting galaxy
scenario is preferred over gravitational lensing or an intervening
spiral galaxy as an explanation for
the origin of the structure (\S~\ref{lensmerger}); model the 
stellar populations in the arcs (\S~\ref{stellarpop} and
\ref{mass}); and look at the link between the dynamics of the
assumed galaxy interaction and the activity in the nucleus
(\S~\ref{plane}). 
In \S~\ref{1517conc} we summarize our results.

The cosmology used throughout this paper is 
$H_0=70$~km~s$^{-1}$~Mpc$^{-1}$, $\Omega_M=0.3$, and $\Omega_{\lambda}=0.7$.

\section{Observations and Data Reduction}
\label{obs}

H1517+656 was imaged with HST on the 17th of June 2000 
using the STIS CCD with the F28$\times$50LP longpass 
filter and the PC2 CCD with the F555W and F814W filters. The STIS 
observations provide very deep, high-resolution imaging for lens 
modelling or morphological analysis, while the PC2 observations provide colour 
information to test the lensing hypothesis and allow modelling of stellar 
populations. Table~\ref{obstab} gives the exposure times for these observations.

\subsection{Data Reduction and Sky Subtraction}\label{datared}

\begin{figure}
\figurenum{2}
\plotone{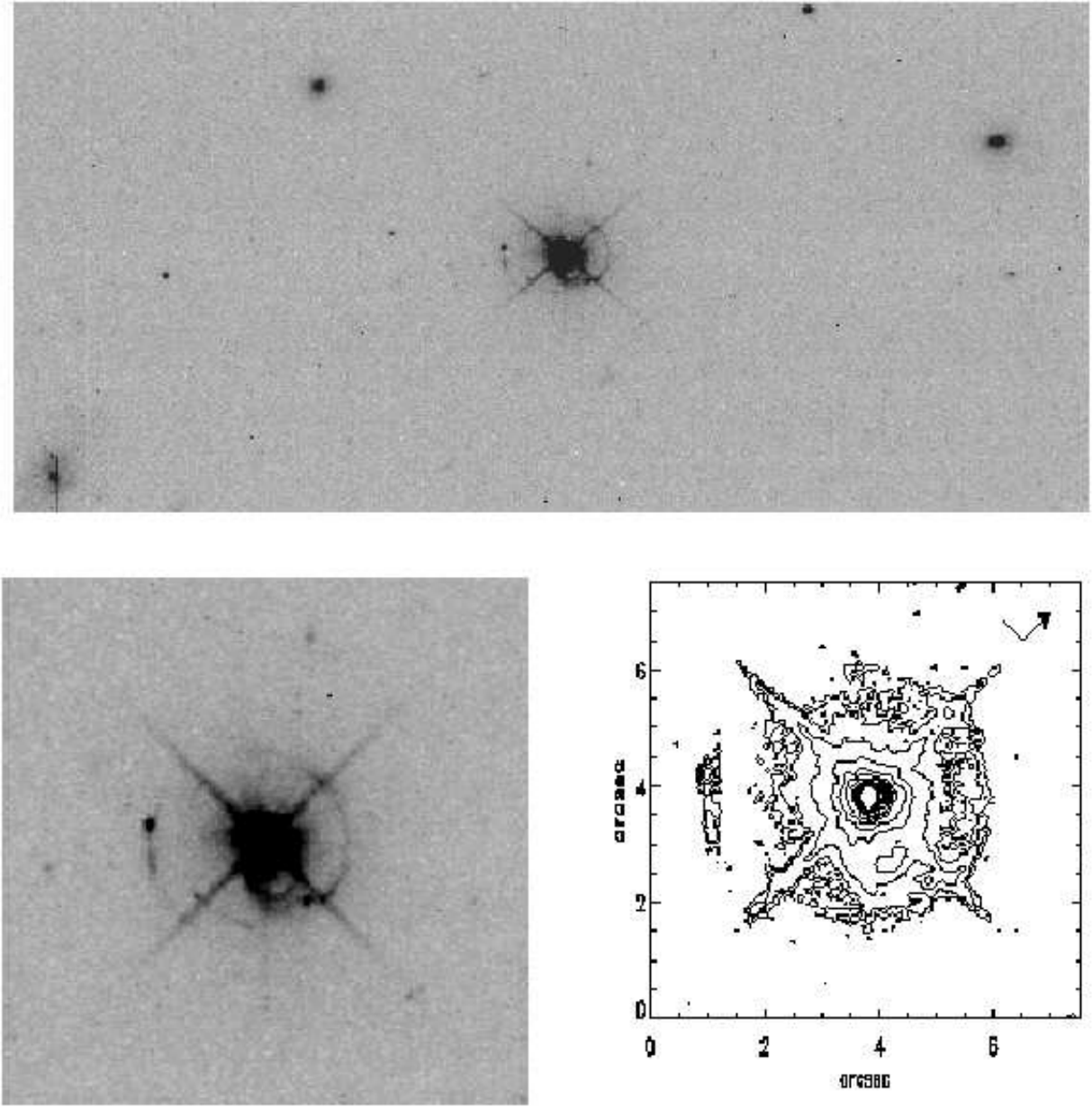}
\caption{
Reduced, sky-subtracted STIS image of H1517+656 in the
F28$\times$50LP filter. The {\it top} image is the full CCD, the 
bottom {\it left} is the 300$\times$300~pixel region surrounding the
target, and the bottom {\it right} is the contour plot of the image,
with the arrow indicating the directions of north and east.
\label{stis1517}
}
\end{figure}

\begin{figure}
\figurenum{3}
\plottwo{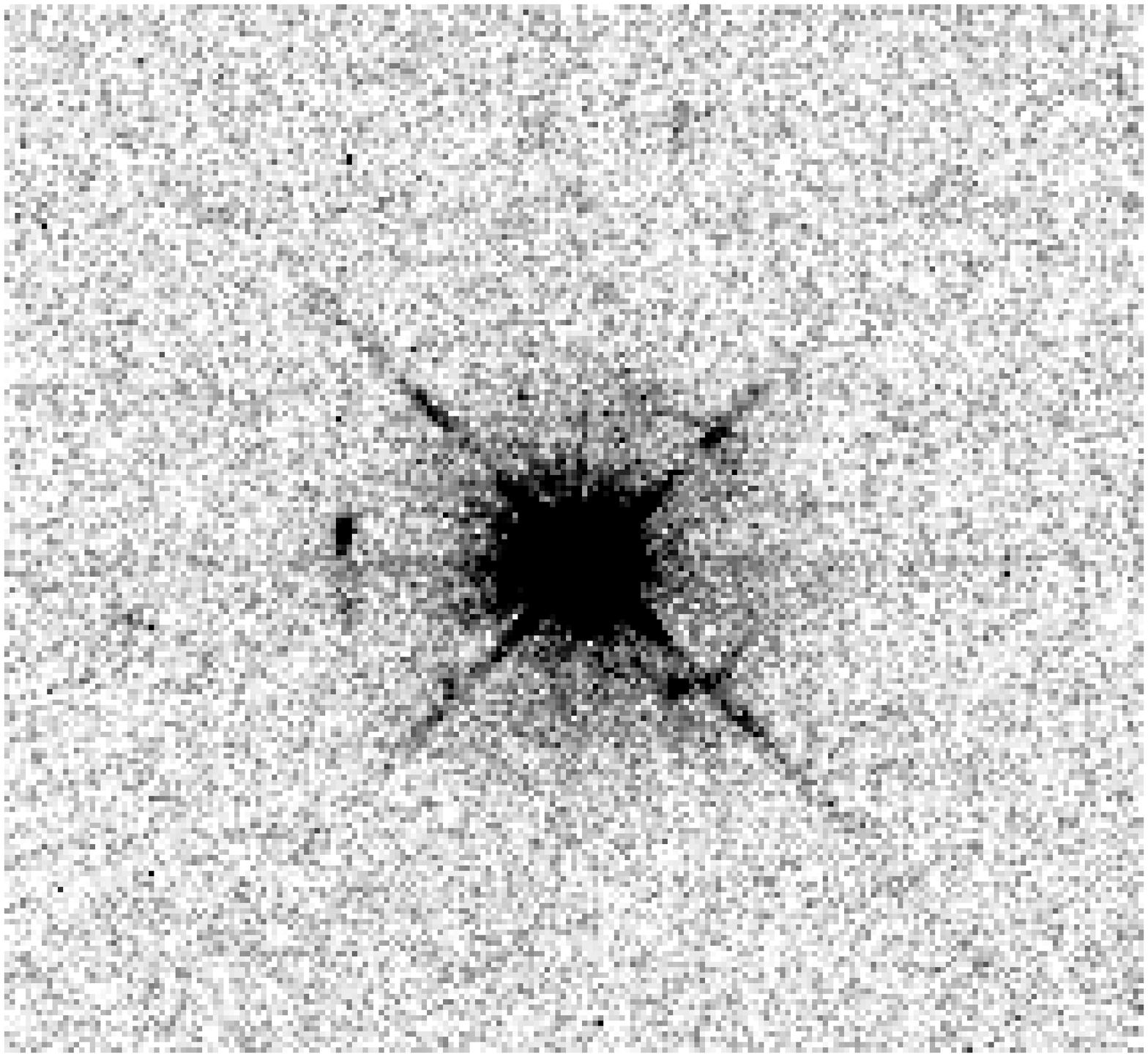}{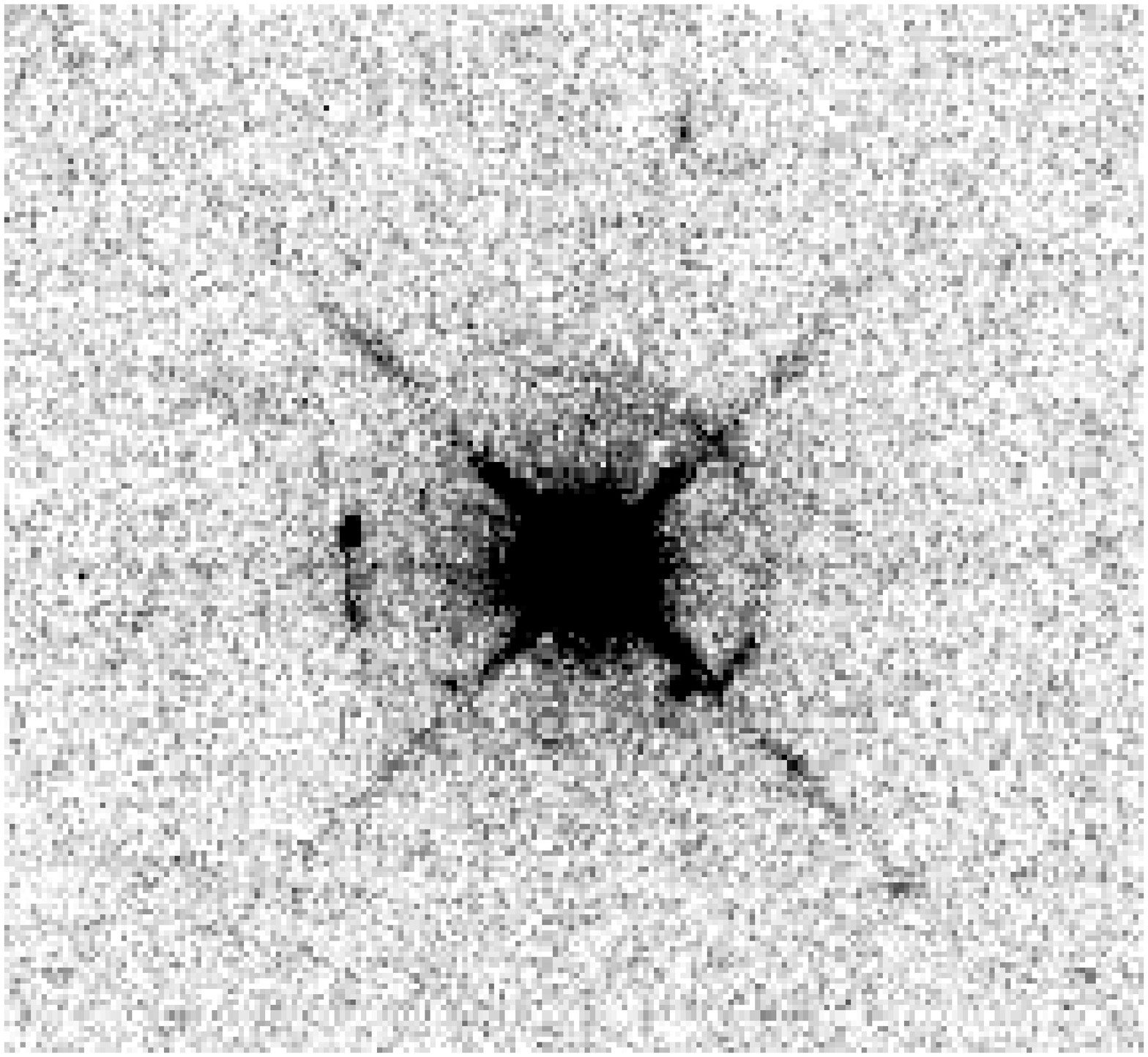}
\caption{
Reduced, sky-subtracted WFPC2 images of H1517+656 in the F555W filter
({\it left}) and the F814W filter ({\it right}). The
300$\times$300 pixel region surrounding the BL~Lac in the PC camera is
shown. Orientation is the same as in Figure~\ref{stis1517}.
\label{wfpc1517}
}
\end{figure}

The data were bias-subtracted, flat-fielded, cosmic ray-rejected and
combined using STIS and WFPC packages in IRAF's STSDAS suite. 
The background was taken to be the median in the region of each
combined imaged that was unaffected by visible sources. Such
sources were masked well beyond their visible extent for determination
of the background.
The error in the background value was calculated from two sources --- the
pixel-to-pixel poissonian shot noise ($\sigma_{poisson}$), and the
large scale variation in the background level ($\sigma_{large~scale}$). 
The latter was taken to be the maximum of: the standard deviation in the
medians of a grid of 20$\times$20 pixel squares, and the standard deviation
in the medians of the four quadrants of the image. The final
error in the subtracted sky value was then: 
$\sigma_{sky} = \sqrt{\sigma_{poisson}^2 + \sigma_{large~scale}^2}$.
Figures~\ref{stis1517} and ~\ref{wfpc1517} show the reduced,
sky-subtracted images.

\subsection{Modelling the Point Spread Function}

While the model Point Spread Function (PSF) generated by the Tiny Tim
\citep{Krist} software is useful in finding the detailed structure of
the PSF, it does not account for the scattered light in the wings
\citep{Scarpa}, or for the well-known 'ghost loop' reflection feature in the STIS
PSF. As the arcs around H1517+656 have very low surface brightnesses,
it was critical that we subract the contribution of the PSF wings
carefully. To do this, we constructed composite PSFs for each 
filter. We modelled the outer regions from several highly over-exposed
images of stars taken from the HST archive, and the central regions
from Tiny Tim models for the PC images, and from unsaturated archival
stellar images for the STIS longpass image. 

Given the potential inaccuracies in the STIS PSF extreme care was
taken to account for its uncertainties. Deviations in the PSF 
arise from spatial variations, time variations, and spectral
variations. The stars that were averaged to make our composite PSF
have different spectral types, were taken at different times, and are
offset from each other spatially on the detector (although are all still 
within 50 pixels of the target). Thus, the deviations observed between these PSFs
should reflect the potential difference between our composite PSF and
our data. The STIS ghost loop varies significantly even over small 
spacial offsets, and with the lack of exactly-positioned archival stars, 
a perfect model of this feature was not possible.

We define the uncertainty in each pixel of the composite PSF model to be
the maximum difference among that pixel value and the corresponding
pixels in the component stellar images. This gave a conservative
uncertainty map that was used in the two-dimensional PSF and host galaxy
fitting. 

\subsection{Host Galaxy Analysis}

The deep STIS image was used to analyse the host galaxy. We first mask
the regions of the arcs and the PSF ghost loop conservatively, including
several pixels beyond their visible extent. We then use $\chi^2$ minimization 
to fit the best two-dimensional model. Models comprised either 
the PSF superimposed with a centered de~Vaucouleurs galaxy model, the
PSF with an exponential disk model, or the PSF alone.

The significance of the fit was determined by comparing $\chi^2$ for
of the best-fit PSF + single component host galaxy model with $\chi^2$
for the best PSF-only fit via the F-test. To claim a host galaxy
detection, we required that the host
galaxy fit be preferred over the PSF-only fit at the 99\% confidence
level. We also required that the host galaxy fit be
preferred over the PSF-only fit when the sky is over-subtracted to the
$1\sigma$ limit of the estimated sky error, at the 95\% confidence level.

A host galaxy appeared to be marginally detected in the initial
fitting process, however it did not satisfy the conditions described
above. The best PSF + de~Vaucouleurs fit improved the PSF-only fit
with marginal significance: 97.5\% by the F-test. When the sky was
over-subtracted, this improvement was reduced to 92\% significance.
Thus, the host galaxy is considered unresolved. The use of more
complex modelling is unjustified given this lack of detection.

We calculated the upper limit on the host galaxy brightness by
finding the most luminous host galaxy that could be added before the
fit became worse than the PSF-only fit at the 99\% confidence
level. We assumed a de~Vaucouleurs profile with an effective radius of
10~kpc. This is higher than the median of 6.35~kpc observed at low
redshift, yielding more conservative limits \citep{Urry}.
The F28$\times$50LP apparent magnitude upper limit was found to
be 19.81~mag, not K-corrected or corrected for extinction. The
absolute R-band upper limit is $-24.6$~mag, assuming $z=0.702$,
corrected for galactic extinction with F28$\times$50LP band
extinction of $A_{LP}=0.068$ (interpolated from results of Schlegel,
Finkbeiner \& Davis 1998), and K-corrected and converted to Cousins R
band assuming the spectrum of a burst model from the 
stellar population synthesis models of \citet{Bruzual} calibrated to
early-type galaxy colours ($B-V=0.96$; Fukugita, Shimasaku \& Ichikawa 1995). 

\begin{figure}
\figurenum{4}
\plotone{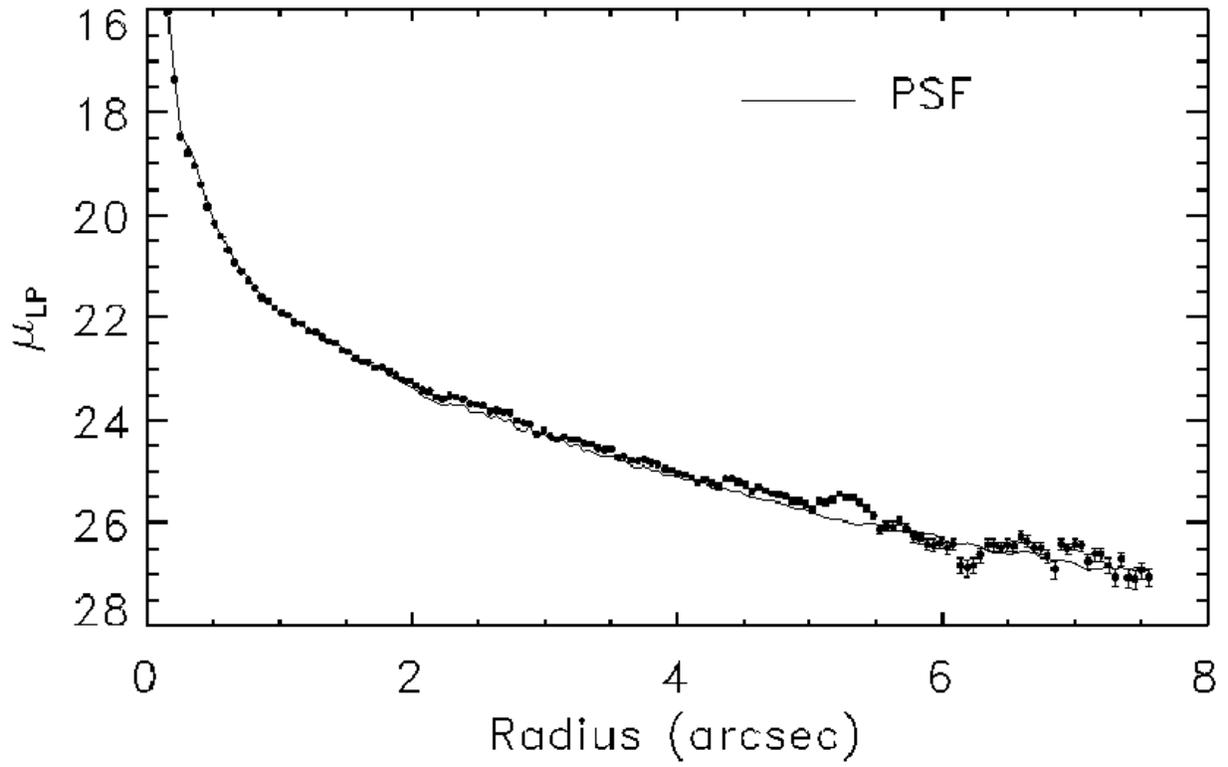}
\caption[Azimuthally averaged profile of H1517+656 with best-fit PSF]{
Azimuthally averaged profile of H1517+656 ({\it data points}), with
the best-fit PSF {\it solid line}.
\label{psf1517}
}
\end{figure}

\begin{figure}
\figurenum{5}
\plottwo{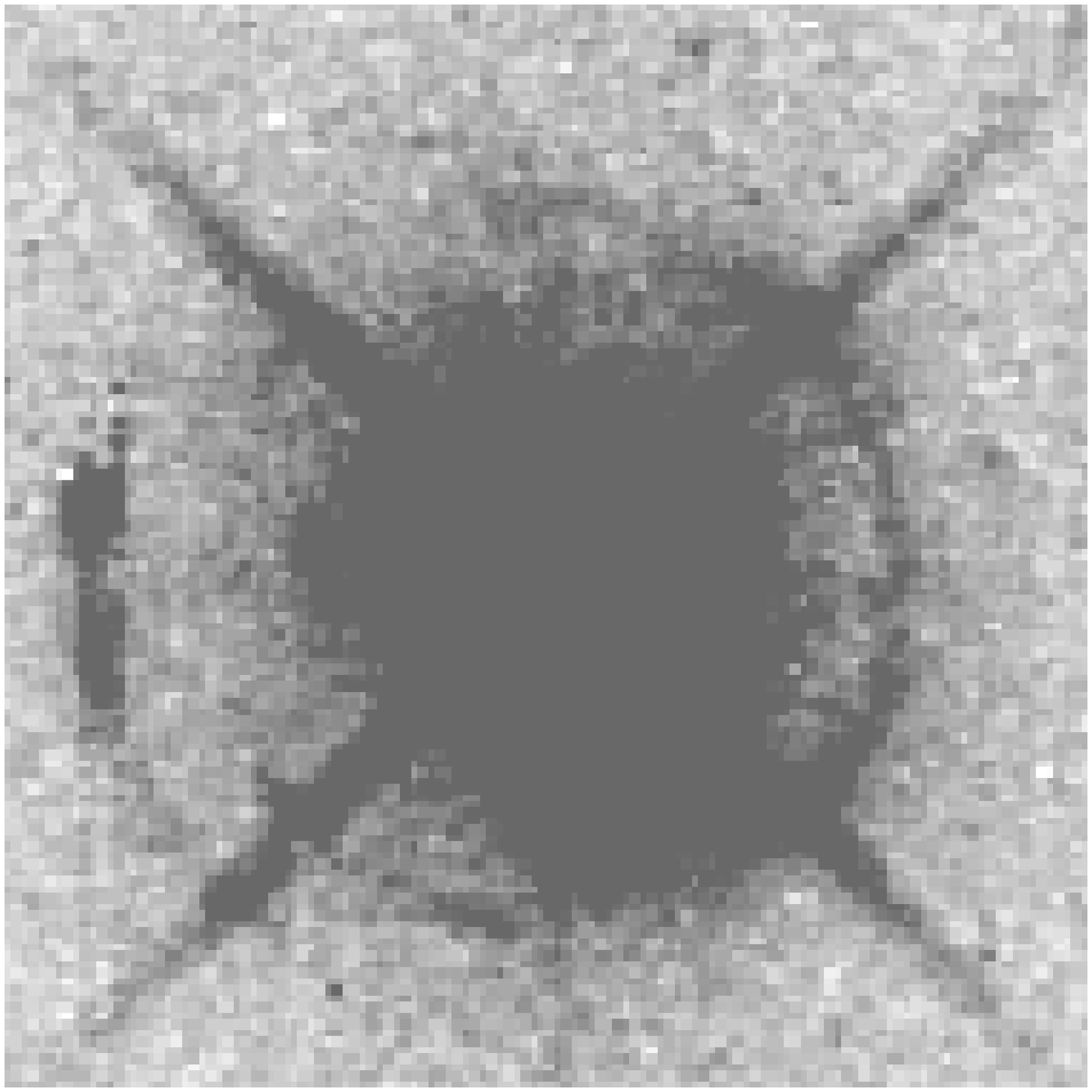}{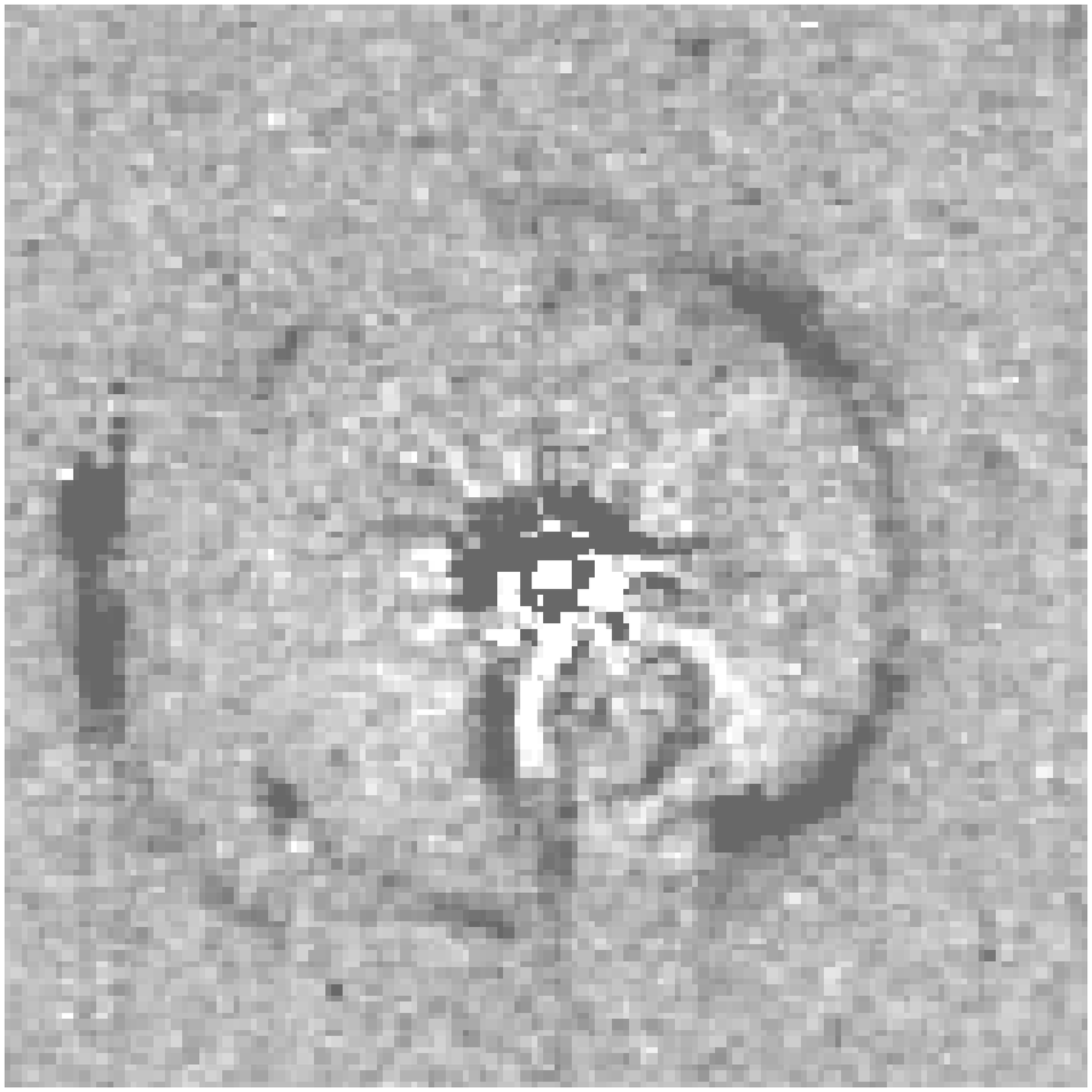}
\caption{
Unsubtracted STIS image of H1517+656 ({\it left}), and 
with the best-fit PSF subtracted ({\it right}). The circular feature
to the lower-right of the nucleus is part of the PSF. This reflection 
feature is highly sensitive to chip position, and so could not be 
subracted perfectly (but was masked in the fitting process). Imperfect
subtraction yields the adjacent bright and dark features seen around
the ghost loop. The absence of such features near the arcs indicates
that the PSF is accurately modelled in these regions.
\label{hostsub}
}
\end{figure}

Figure~\ref{psf1517} shows the azimuthally averaged profile of
1517+656 with the best-fit PSF. Figure \ref{hostsub} shows the
original STIS image, and the image with the normalized PSF subracted. 
The PSF-subtracted image does not reveal obvious excess light that
might be associated with the host galaxy. The arcs, however, are
clearly not part of the PSF. Significant additional structure is
revealed in the subtracted image.

\subsection{Photometry}

\begin{figure}
\figurenum{6}
\plotone{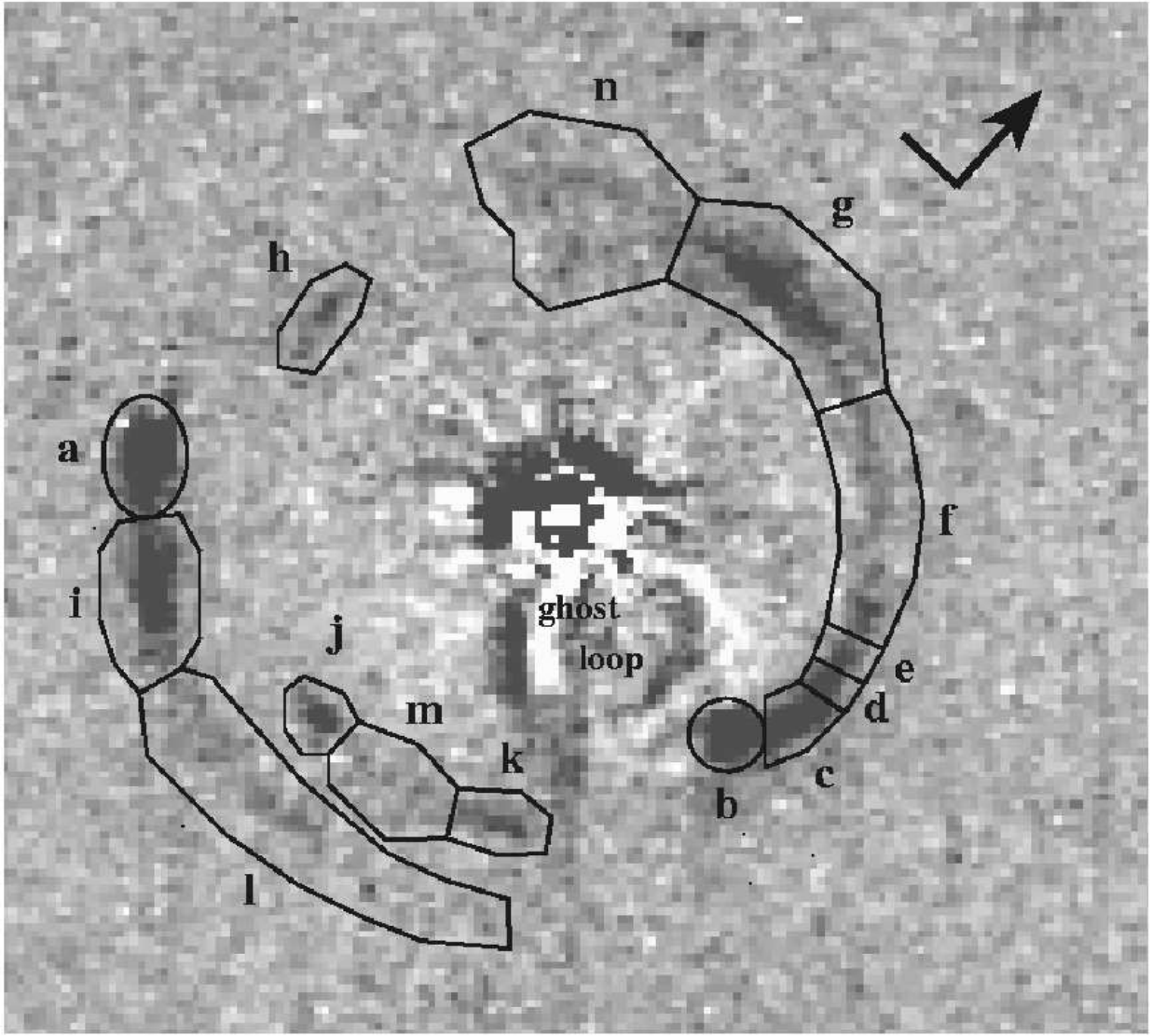}
\caption{
Labels assigned to the structure components around H1517+656. Lines
indicate apertures used for determination of magnitudes and
average surface brightnesses. North and east are indicated by arrow
head and tail respectively.
\label{1517label}
}
\end{figure}

After subtracting the best-fit PSF from the image, aperture photometry
was performed on the arcs and knots. The results are in
Table~\ref{arcphot}. The labelling of the arcs and the apertures used
are shown in Figure~\ref{1517label}. Segments are divided according to 
both spatial separation and color differences. Apparent magnitudes are given for
the knots {\it a} and {\it b}, and the errors in these values are  
$\sim\pm$0.02~mag, dominated by uncertainty in the background subtraction,
and affected also the uncertainty in the PSF subtraction.
The surface brightness of the arcs vary strongly in the radial
direction, and so their average values depend on the
width of their apertures. The uncertainty in these surface brightnesses
is $\sim~\pm0.15$~mag, and includes both the uncertainty in the
background and PSF subtraction, the variation in surface brightness
across the extend of each arc segment.

Table~\ref{arcphot} also gives the $V-I$ colors of each
component. These are derived from $F555W-F814W$ color using $V-F555W=0$ 
and $I-F814W = 1.22$ (WFPC2 Instrument Handbook). The error in the
conversion from HST magnitudes to V and I is of order 1\%. The
variation in color across the region of each arc segment is smaller
than the variation in surface brightness, and so the overall
uncertainty in the arc colors is  $\sim~\pm$0.1~mag. The uncertainty
in the knot colors is $\sim0.03$~mag.

\section{Analysis and Discussion}
\label{1517disc}

\subsection{Gravitational Lens, Intervening Spiral, or Colliding Galaxies?}
\label{lensmerger}

The additional detail revealed in the STIS image all but completely rules out
gravitational lensing as the explanation of the arcs of
H1517+656. Certainly the structure could not result from a single
extended source --- the two longer arcs have distinctly different
radii of curvature (see \S~\ref{plane}). It is conceivable that a
complex configuration of extended structures may be able to produce
the observed structure. To test this, the image was analysed using
Lensview \citep{Wayth} --- a program for deconvolving extended
gravitational lenses based on the LensMEM algorithm of \citet{Wallington}

No background source configuration could be found that
yielded the observed light distribution for any reasonable mass
distribution of the lens system. The primary difficulty was in
matching up opposing lens images, for example:

\begin{itemize}

\item
Unless the mass distribution of the
lens is very complex, we expect every image to have a counter image
that may differ in its transverse stretching, but should not differ
in its radial stretching. The fact that component {\it b} is closer to 
the nucleus than the rest of northwest arc described by components 
{\it c} to {\it g} means that this structure is not the opposing 
image of the south filament ({\it a} to {\it l}), as this latter 
filament does not have a corresponding radial kink.

\item
If component {\it b} does not correspond to component {\it a}, then
the only other opposing structure is component {\it h}. These two
components are almost equidistant from the (assumed) center of mass, and so
should be of similar brightness to each other and exhibit more transverse
sheer.

\item
The structure at {\it n} is quite extended radially (although
difficult to see in the printed image due to its extremely low surface
brightness). The only possible counter-images for this component are
{\it m}--{\it k} or {\it l}, and neither of these have similar radial
extent. 

\end{itemize}

Additionally, the spectral information gained with the F555W and F814W
filters (Table~\ref{arcphot}) shows substantial color variation
across the structures. In particular, the knots {\it a} and {\it b}
are much bluer than many of the arc segments ({\it c, e, f, h} \& {\it i}). 
Gravitational lens images from the same source have the same intrinsic
spectrum, and hence the same intrinsic colors. Even if there were a sensible
lens geometry by which the knots and arcs could be credibly associated
with each other, they have the largest color differences, and so
cannot be lensed images of the same source component.
The observed difference in  
color could be due to differential reddening of some of the images 
due to dust in the lensing galaxy, but the chance alignment of
the all bright lens images with dust-free regions, and the extended
arcs with dust-thick regions is too coincidental to be credible.

The increased sensitivity granted by the STIS image also allows us to 
completely rule out a  spiral galaxy as the explanation of the
structure --- either in the foreground or hosting the
BL~Lac. Although more arc fragments were revealed, these are clearly
fragmented. They are not connected in spiral arms, nor do they extend
to the central regions. 

If the arcs are not gravitational lens images or an intervening spiral
galaxy, then they are likely to be physically associated with the BL
Lac object, and so probably result from a galaxy interaction.
A number of types of interacting systems need to be considered; 
we may be looking at a ring galaxy, the limb-brightened shells of a
shell galaxy, or the accreting material from a close interaction. 
Although the diameter of the ring structure is similar to that of some 
ring galaxies (e.g. The Cartwheel, with its 33~kpc radius ring), the
multiple levels of concentric arcs (at least two different radii) in
1517+656 make this explanation
unlikely. Limb-brightened shells are similarly unlikely: the presence
of three bright knots (including component {\it j}) coincidentally
situated on the very limbs of these shells is quite improbable.
The most likely explanation is that the arcs are the accreting debris
from a merger event or a close interaction in which material was
stripped from the passing galaxy. Dynamical friction with the host
galaxy may then have resulted in the debris settling into the observed 
near-circular arcs.

\subsection{Modelling the Stellar Population}
\label{stellarpop}

Assuming that the observed structure does result from a 
galaxy--galaxy interaction, bursts of recent star formation induced by
this interaction should have an observable effect on the spectral energy 
distribution (SED). Indeed, the colors measured (Table~\ref{arcphot}) are of 
very blue populations. We can use stellar population
synthesis models to reproduce the $F555W-F814W$ colors of the
structure components, and so place contraints on their star formation
history. 

We used the GISSEL evolutionary synthesis code of \citet{Bruzual}, with the 
{\em Padova 2000} library of isochrones \citep{Girardi},
and the BaSeL standard stellar library \citep{Lejeune}, 
extended to non-solar metallicities \citep{Westera}.
The stellar populations were modelled as simple bursts of star
formation. Synthetic SEDs were generated for a grid of population
ages, from $1.5 \times 10^{5}$~years to $1.8\times 10^{10}$~years, and
for metallicities ranging from 0.02 solar to 1.5 solar. We probe to very 
low metallicities to determine whether metallicity alone can explain the blue 
colors, or whether young ages, as expected from an interacting system, are 
indeed required. We investigate a higher-than-solar metallicity
to demonstrate that small changes in this direction have little effect
on the ages. Significantly higher metallicities will result in even
younger model populations fitting  
the data. The SEDs resulting from these models
were then redshifted over the range $0 \le z \le 5$ and convolved with
the HST transmission curves. In this way we obtained color as a
function of age, redshift and metallicity.

\begin{figure}
\figurenum{7a}
\plotone{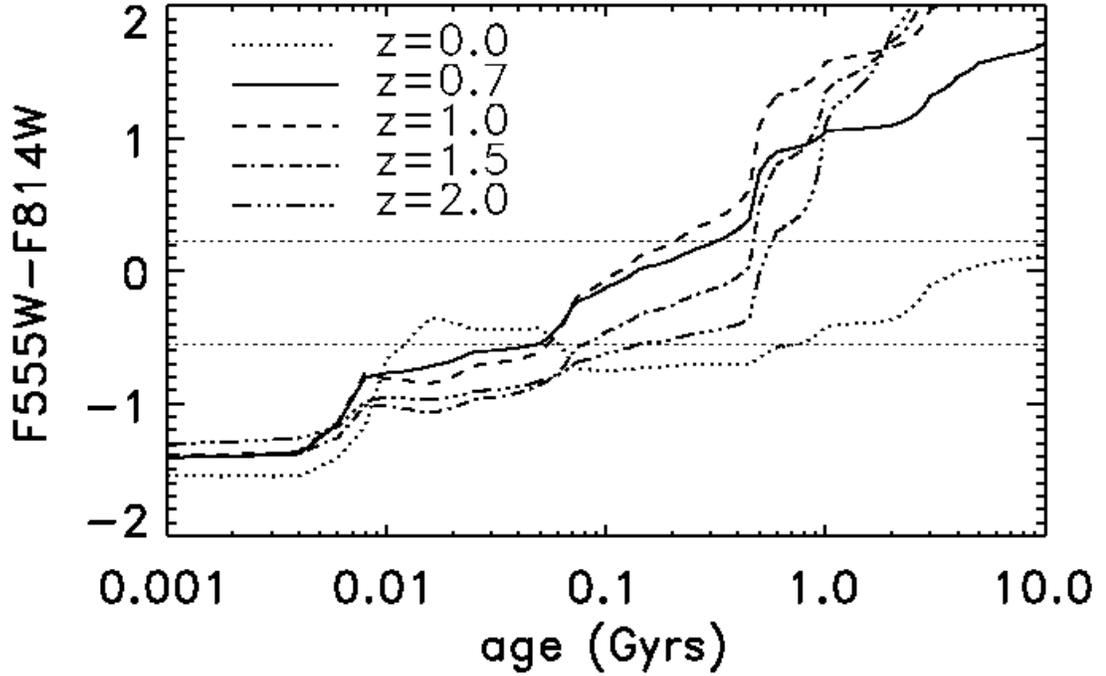}
\caption{
$F555W-F814W$ colors of a model solar metallicity burst population as a function of age
for a range of redshifts. For redshifts $z \gtrsim 0.7$, color
becomes steadily more red with increasing age. The {\it horizontal dotted
lines} show the $F555W-F814W$ colors of the reddest and bluest of the
structures surrounding H1517+656, indicating a range of ages from
$\sim 0.02$~Gyrs to $\sim 0.6$~Gyrs for all possible redshifts ($z \ge 0.702$). 
\label{colorageredshift}
}
\end{figure}

\begin{figure}
\figurenum{7b}
\plotone{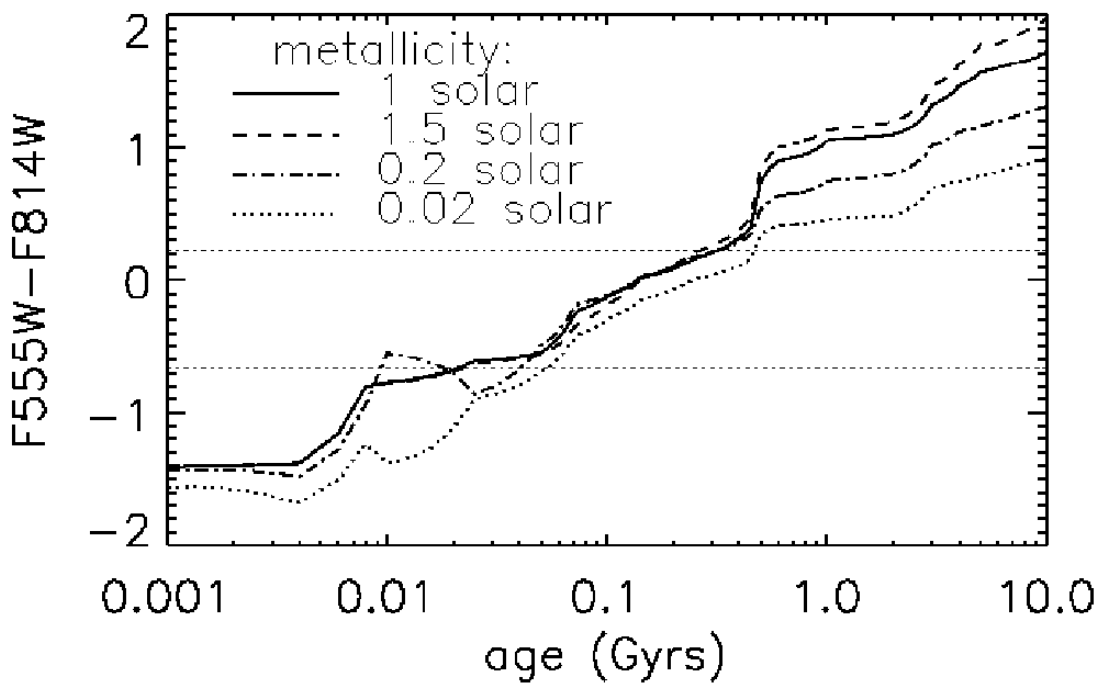}
\caption{
$F555W-F814W$ colors of a burst population at $z=0.702$ as a function
of age for a range of metallicities. It can be seen that metallicity
has only a small effect on the age range of the structures surrounding
1517+656 (the {\it horizontal dotted lines} show the bluest and
reddest of these colors), implying $age \lesssim 0.03$~Gyr and
$\lesssim 0.3$~Gyrs respectively for most metallicites ($\lesssim
0.045$~Gyrs and $\lesssim0.45$~Gyrs for an extremely low metallicity,
$0.02$~solar). 
\label{coloragemetal}
}
\end{figure}

The resulting function is, as expected, sensitive to age, but
less so to redshift and metallicity. 
Figure~\ref{colorageredshift} shows $F555W-F814W$ color as a
function of age for redshifts $z=0$ to 2, for a solar metallicity
population. Color increases roughly
linearly with the log of age over the color range of interest, with
the age of the structure falling between 0.04 and 0.57 Gyrs for all
possible redshifts ($z \ge 0.702$). Figure~\ref{coloragemetal} shows 
$F555W-F814W$ color as a function of age for a range of
metallicities, assuming $z=0.702$. Only the lowest metallicity
modelled (0.01~solar) has an appreciable affect on the maximum age,
increasing it by $\sim 50\%$. 

\begin{figure}
\figurenum{8}
\begin{minipage}{0.45\linewidth}
\plotone{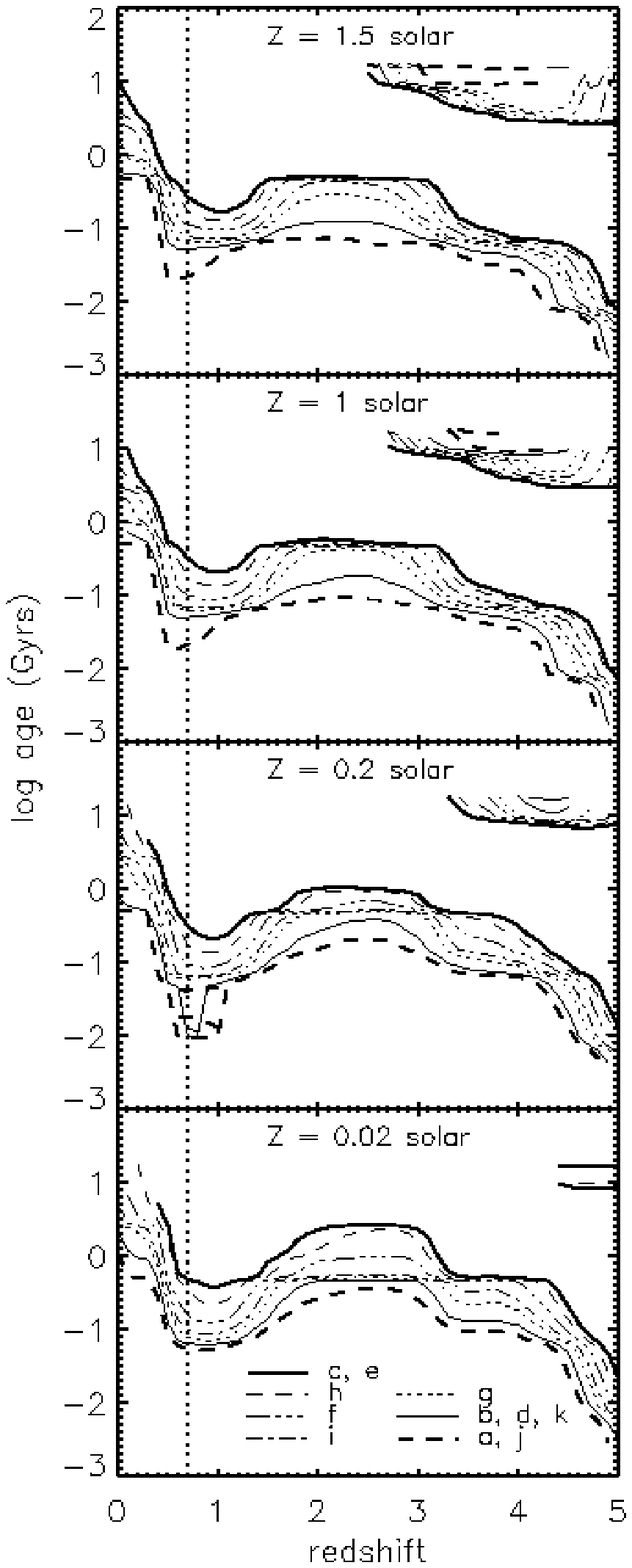}
\end{minipage}
\hspace{1cm}
\begin{minipage}{0.4\linewidth}
\caption{\small
Age as a function of redshift for a burst population with the 
$F555W-F814W$ colors of the structure observed around H1517+656, for 
four metallicities (from highest to lowest: $Z=0.03$, 0.019 (solar),
0.004 and 0.0004). 
For metallicity $\ge 0.2$~solar at $z=0.702$ ({\it vertical dotted
line}), the majority of the components have age $<$ 0.1 Gyrs, and
are all younger than 0.6~Gyrs, regardless of metallicity. 
At some redshifts, particularly the higher ones, multiple burst ages
can reproduce the colors observed in the structures. The degenerate
solutions at $z=0.702$, however, all fall within a small range of
ages.
\vspace*{10cm}
\label{agez}
}
\end{minipage}
\end{figure}

Figure~\ref{agez} shows age of burst
population with the $F555W-F814W$ colors of the structure
surrounding H1517+656, as a function of redshift for four different
metallicities. For from 1.5 to 0.2~solar, most of the components 
have $age < 0.1$~Gyrs for $0.702 \le z \lesssim 1.1$, which covers the
range of likely redshifts of the source. All components have 
$age < 1$~Gyrs for $0.702 \le z \lesssim 2.5$. 
For extremely low metallicity (0.02~solar), the ages are still low, 
but are $1.2\times$ to $2.5\times$ the solar metallicity ages.
The brightest, most compact structures appear the youngest. At
$z=0.702$, the knots {\it a} \& {\it j} have a solar metallicity 
burst population age of 21~Myrs, and {\it b} \& {\it d} have a burst 
age of 42~Myrs. 

Particularly at high redshift ($z > 2.5$), burst populations with
quite different ages may produce the same $F555W-F814W$
colors. However, at
lower redshift, for the colors observed in the structures, any
degenerate solutions all lie within a narrow range of ages.

Table~\ref{arcages} summarizes the ages found for the different
components, assuming $z=0.702$. The errors given in this table
are derived from the 1$\sigma$ confidence limits in the $F555W-F814W$
colors. In some cases, the upper or lower limit of these colors had 
multiple solutions in the age--redshift relation. In these cases, the
errors presented represent the full span of these degenerate
solutions. Components {\it l, m} \& {\it n} are not included as they
are not detected in the F555W and F814W filters.

\subsection{Dynamical Age Estimate}
\label{dynamics}

The ages derived above assume that all of the luminous mass in the
arcs results from recent bursts of star formation. The true scenario is
certain to be more complicated, and so these ages don't directly 
give us the time since the interaction began. 
An independent estimate of the minimum time since the merger began 
can be made by calculating the time required for the material 
to spread over the circular extent of the arcs --- 
$\sim$15~kpc in radius if $z=0.702$ --- assuming orbital velocities.

Assuming that the host galaxy has a luminosity and morphology similar 
to low-redshift BL~Lac host galaxies --- $M_R=-22.85$~mag and a 
bulge-dominated profile with $r_e=6.35$~kpc \citep{Urry}
--- and a mass-to-light ratio of $M/L_V=25$~M$_{\odot}$/L$_{\odot}$ 
\citep{Loewenstein}, we obtain
a mass of $\sim2.3\times10^{11}$~M$_{\odot}$ within a 15~kpc radius. 
This gives a Keplerian velocity of $\sim400$~km/s for the arcs. At this 
velocity, the material would have taken $\sim0.2$~Gyrs to spread 
around the host galaxy. A shorter timescale is
possible as the accreting material is expected to have been slowed 
by dynamical friction with the host galaxy. However, if this dynamical
friction is also to explain the circularity of the arcs, then we 
expect that the debris has been settling for at least one orbit of 
of the host galaxy.

\subsection{The Mass of the Arcs}
\label{mass}

Table~\ref{arcmasses} shows the luminous mass of a burst population required to
produce the flux observed in the STIS image assuming z=0.702, solar metallicity, and
the burst ages given in Table~\ref{arcages}. For components {\it l, m}
  \& {\it n} we assume the same $F555W-F814W$ color as component {\it f},
the most comparable extended, diffuse arc segment. The
total mass, assuming that all of the observed luminosity comes from
the recent starburst, is 2.17$\times 10^{9}$M$_{\odot}$, for solar 
metallicity. This is an upper limit on the true burst population mass,
as there is certain to be an underlying, older population. 

If we assume this older population has a spectrum similar to a spiral galaxy
(more likely than an early-type spectrum, as a significant amount of
gas must have been present to fuel the observed star formation), 
then we can calculate the actual mass of the burst component required
to reproduce the observed colors for a range of burst ages. 

The underlying spectrum is derived by calibrating the GISSEL models to
the $F555W-F814W$ colors of a Hubble type Scd galaxy
($F555W-F814W=1.02$; Fukugita, Shimasaku \& Ichikawa 1995). Table~\ref{arcmasses} gives the
derived masses of both populations for 
burst ages of 0.001, 0.01 and 0.02 Gyrs. We limit the burst age to
0.02 Gyrs at the oldest because the colors of older burst populations
are redder than many of the arc components, and to 0.001 Gyrs at the
youngest because burst populations younger than this are not
significantly bluer.

The total luminous mass of the arcs is remarkably high:
at least $\sim 1.5\times 10^{9}$~M$_{\odot}$, and perhaps up to 
$3.7\times 10^{9}$~M$_{\odot}$.

\subsection{The Plane of the Collision}
\label{plane}

If the observed arcs are really the remnants of a collision, their
apparent circularity is remarkable. The northwest arc joining components {\it c}
to {\it g}, in particular,
appears to have highly circular curvature to the eye, while
the southeast ({\it a} to {\it l}) and south arcs ({\it j} to {\it k}) show an
increase in radius of curvature with increased distance from the
assumed center of mass (the nucleus of the BL~Lac).

The observed circularity could be the result of the chance projection
of much more elliptical arc curvatures at a larger angle to the plane
of the sky. However, if the arcs are in the same plane, this would
require that the semi-major axes of the 
arcs all be aligned. While we expect the plane of each arc to be
linked to the plane of the interaction, and hence to each other, the
same is not true of their semi-major axes. The alignment of each 
arc axis is defined by the direction in which the stellar
material was moving when it was thrown off during merger, or the
direction in which its parent galaxy was moving when it was stripped. 
The northwest and southeast arcs are clearly distinct streams of material,
well-separated and with different radii of curvature. It is expected
that for any close interaction, as this one must have been, the
trajectory of the incoming galaxy would change considerably over the
course of the encounter. Thus there is no reason to expect two
well-separated streams of accreting material to have 
high ellipticities {\it and} aligned axes. It could happen,
but a naturally circular geometry, perhaps resulting from friction
between the accreting material and the halo of the host galaxy, seems
the simpler explanation. 

Thus, assuming that the arcs are intrinsically circular, we can
constrain their angle to the sky by fitting ellipses to each, and
finding their maximum possible ellipticities.
Ellipses were fitted to each of the three prominent arcs separately,
allowing ellipticity, semi-major axis and 
position angle to vary freely, while fixing the ellipse center to the
BL~Lac nucleus. $\chi^2$ minimization was used, with each point on the
arcs weighted inversely to the half-light width of the arc at that
point. Table~\ref{ellipsefits} shows radii of the three most 
prominent arcs along with the results of the ellipse fitting.
If the arcs are intrinsically circular, then their maximum angle to
the sky is $\sim20^{\circ}$ for the well-constrained northwest arc and 
$\sim25^{\circ}$ for the southeast arc, determined independently to
 each other (see Table~\ref{ellipsefits}). 

This apparent alignment of the arcs to the plane of the
sky is intriguing. The currently accepted model of
the BL~Lac central engine involves a relativistic jet with a small
angle to the line of sight~\citep{BlandfordR}, a model supported by 
their high polarisations, intra-day variability, asymmetries in
brightness and polarisation, and unphysical brightness temperatures. 
If the jets of these objects are powered by the magnetic fields of
rapidly-rotating black holes, then we expect that the angle of
these jets is defined by the direction of spin of the black hole. 
If the black holes in radio-loud AGNs are spun up by the transfer of
angular momentum during mergers, then we also expect there to be a
correlation between the plane of the interaction and the angle of the
resultant jet. 

With H1517+656 we may be able to link the generation of
an AGN jet with a merger event. This provides evidence that the 
physical processes that result in AGN jets are dependent on the 
plane of this interaction, and hence on the transfer of angular 
momentum in the interaction.

Regardless, this link between the geometry of the merger and the
emission of the central engine indicates that the AGN activity
observed in this source may have been triggered, or at least strongly
influenced by the current interaction. 
The ages of the stellar populations are consistent with this
hypothesis, as is the size of the arc structure if we account for 
the infall time of material into the AGN core.
The extreme emission properties of this BL~Lac, from
radio to X-ray frequencies, also suggest that this source is in an
unusual phase in its life. 

\section{Conclusions}
\label{1517conc}

The presence of unusual, ring-like structure surrounding the BL~Lac object
1517+656 has been investigated with HST imaging. 
Observations in the STIS longpass filter have allowed us to
better constrain the upper limit on the host galaxy's brightness to
$M_R > -25.23$, and to resolve the arcs in much greater detail.
WFPC2 F555W and F814W observations have given us spectral information
on the structure components.

Analysing the structure of the arcs with lens deconvolution software,
we were able to determine that they are highly unlikely to be
gravitational lens images from morphological arguments alone.
However, the strongest argument against gravitational lensing is that the
$F555W-F814W$ colors show strong variation between sections of the arcs
that should correspond to the same source components.

Rather, these structures are likely to result from 
a close galaxy interaction or merger, and consist of stellar material and
gas accreting onto the BL~Lac host galaxy. This hypothesis is
supported by the extremely blue colors observed in the arcs and knots.
The $F555W-F814W$ colors indicate very young stellar populations, with
ages less than 0.02 Gyrs in parts of the structure. 
From stellar population models, we a derive a luminous mass
of between $1.5\times 10^{10}$~M$_{\odot}$ and 
$3.7\times 10^{10}$~M$_{\odot}$ for the structure. The high mass 
and the starburst colors of the accreting material indicate that the 
interaction was a violent one. 

The apparent circularity of the arcs indicates that the plane of the
collision, and hence accretion, is very close to the plane of the
sky --- within $20^{\circ}$ if the arcs are approximately
circular. This is intriguing, as it would mean that the plane of the 
interaction is perpendicular to the jet, which
is thought to be aligned with the line of sight in BL~Lacs.

The link between the plane of the merger and the angle of the jet
suggests that activity in H1517+656 may have been triggered, or at
least strongly influenced, by this interaction. If so, then this is a
direct example of the transfer of angular momentum in a merger resulting
in the generation of a radio source. The extreme emission
properties of the core supports the hypothesis that H1517+656
is a radio-loud AGN in its birth throes. 


\acknowledgments

\clearpage

\begin{deluxetable}{lll}
\tablecaption{
\label{obstab}
}
\tablewidth{0pt}
\startdata
\hline
\hline
Detector&Filter&Exposure Time (s)\\
\hline
STIS&F28$\times$50LP&2250\\
PC2 &F555W&2460\\
PC2 &F814W&3050\\
\enddata
\end{deluxetable}

\clearpage

\begin{deluxetable}{ccccc}
\tablewidth{0pt}
\tablecaption{Apparent magnitudes and colors of structure components\label{arcphot}}
\startdata
\hline
\hline
    &\multicolumn{3}{l}{Apparent Magnitude:}&\\
Knot&F28$\times$50LP&F555&F814W&V--I\\
\hline
a & 23.62 &  23.37 &  24.03 & 0.56\\
b & 24.19 &  23.73 &  24.30 & 0.65\\
\hline                           
   &\multicolumn{3}{l}{Surface Brightness:}&\\
Arc&F28$\times$50LP&F555&F814W&V--I\\
\hline
c & 22.40 &  22.72  &  22.50  & 1.44\\
d & 22.44 &  22.25  &  22.80  & 0.67\\
e & 23.11 &  23.06  &  22.88  & 1.40\\
f & 24.13 &  23.77  &  23.90  & 1.09\\
g & 22.46 &  22.80  &  23.14  & 0.88\\
h & 23.99 &  23.80  &  23.74  & 1.28\\
i & 21.00 &  22.83  &  23.06  & 0.99\\
j & 23.24 &  22.49  &  22.98  & 0.73\\
k & 21.32 &  22.99  &  24.12  & 0.09\\
l & 24.34 & \nodata & \nodata &  \nodata\\
m & 24.45 & \nodata & \nodata &  \nodata\\
n & 24.41 & \nodata & \nodata &  \nodata\\
\enddata
\tablecomments{
The errors in the apparent magnitudes of the knots are
$\sim\pm0.02$~mag and in the surface brightnesses of the arcs are
$\sim\pm0.15$~mag (includes variation in surface brightness across
each arc region). The errors in the $V-I$ colors of the knots are 
$\sim\pm0.03$ and in the arcs are $\sim\pm0.1$. This latter error is
lower than the individual surface brightness errors because the
variation in color across the region of each arc segment is less than
the variation in surface brightness.
Components {\it l, m \&n} were only
resolved in the STIS image.
}
\end{deluxetable}


\clearpage

\begin{deluxetable}{cccccc}
\tablewidth{0pt}
\tablecaption{Ages of the structures surrounding H1517+656, assuming
  single burst population and $z=0.702$\label{arcages}}
\small
\startdata
\hline
\hline
Comp-&F555W&\multicolumn{4}{c}{Age (Myrs), for metallicity:}\\
onent&~$-$F814W&1.5 solar&1 solar&0.2 solar&0.02 solar\\
\tableline
a&-0.66&\phn20  $^{+15}_{-5 }$   &\phn20  $^{+10}_{-5 }$  &   40 $^{  +5}_{-30}$ &    55 $^{  +5}_{-5  }$\\[0.15cm]
b&-0.57&\phn45  $^{+10}_{-20}$   &\phn40  $^{+10}_{-20}$  &\phn46 $^{  +5}_{-35}$ &\phn60 $^{  +5}_{-5  }$\\[0.15cm]
c& 0.22&   265  $^{+90}_{-55}$   &   315  $^{+90}_{-90}$  &   315 $^{+120}_{-95}$ &   470 $^{ +25}_{-55 }$\\[0.15cm]
d&-0.55&\phn50  $^{+10}_{-30}$   &\phn50  $^{+10}_{-25}$  &\phn45 $^{  +5}_{-40}$ &\phn65 $^{  +5}_{-10 }$\\[0.15cm]
e& 0.18&   240  $^{+80}_{-50}$   &   275  $^{+95}_{-75}$  &   260 $^{+135}_{-75}$ &   460 $^{ +25}_{-95 }$\\[0.15cm]
f&-0.14&   110  $^{+20}_{-20}$   &\phn95  $^{+30}_{-25}$  &\phn90 $^{ +35}_{-25}$ &   150 $^{+745}_{-30 }$\\[0.15cm]
g&-0.41&\phn65  $^{+10}_{-10}$   &\phn60  $^{ +5}_{-10}$  &\phn60 $^{  +5}_{-10}$ &\phn75 $^{ +25}_{-10 }$\\[0.15cm]
h&0.066&   175  $^{+55}_{-40}$   &   185  $^{+70}_{-55}$  &   175 $^{ +70}_{-50}$ &   340 $^{+110}_{-115}$\\[0.15cm]
i&-0.31&\phn75  $^{+20}_{-10}$   &\phn70  $^{+10}_{-5 }$  &\phn65 $^{  +5}_{-5 }$ &\phn95 $^{ +30}_{-25 }$\\[0.15cm]
j&-0.66&\phn20  $^{+25}_{-10}$   &\phn20  $^{+25}_{-10}$  &\phn40 $^{  +5}_{-30}$ &\phn55 $^{ +10}_{-10 }$\\[0.15cm]
k&-0.57&\phn45  $^{+15}_{-25}$   &\phn40  $^{+15}_{-20}$  &\phn45 $^{  +5}_{-35}$ &\phn60 $^{  +5}_{-10 }$\\
\enddata
\end{deluxetable}

\clearpage

\begin{table}
\centering
\caption{
Estimate of stellar population masses in arcs surrounding H1517+656. 
\label{arcmasses}
}
\small
\begin{tabular}{|c|r|rrrr|rrrr|rrrr|}
\hline
Comp-&burst only&&\multicolumn{2}{r}{~~Sb ~~+~~ 0.001}&&&\multicolumn{2}{r}{~Sb ~~+~~ 0.01}&&&\multicolumn{2}{r}{~Sb ~~+~~ 0.02}&\\
onent&($10^6~M_{\odot}$)&&\multicolumn{2}{c}{($10^6~M_{\odot}$)}&&&\multicolumn{2}{c}{($10^6~M_{\odot}$)}&&&\multicolumn{2}{c}{($10^6~M_{\odot}$)}&\\
\hline
a    & 436.0~~~  && 6385 & 37.9 &&& 3550 & 61.6   &&& 3100 & 127.1 &\\
b    &  16.6~~~  && 2015 &  2.4 &&& 1830 &  4.0   &&& 1810 &  82.5 &\\
c    &  16.9~~~  && 1095 &  9.1 &&&  411 & 14.7   &&&  327 &  30.7 &\\
d    & 155.0~~~  &&  571 &  4.7 &&&  215 &  7.7   &&&  171 &  16.0 &\\
e    & 329.0~~~  && 3950 & 12.3 &&& 3030 & 19.8   &&& 2915 &  41.2 &\\
f    &  76.8~~~  && 1360 & 12.1 &&&  452 & 19.6   &&&  340 &  40.7 &\\
g    & 364.0~~~  && 5225 & 24.5 &&& 3385 & 39.7   &&& 3160 &  82.5 &\\
h    & 193.0~~~  && 1755 &  3.3 &&& 1505 &  5.4   &&& 1475 &  11.2 &\\
i    &  34.8~~~  &&  853 &  9.7 &&&  124 & 15.7   &&&   34 &  32.5 &\\
j    & 281.0~~~  && 6895 & 78.5 &&& 1005 & 27.1   &&&  277 & 263.3 &\\
k    & 277.0~~~  && 4735 & 41.4 &&& 1635 & 67.2   &&& 1250 & 139.0 &\\
l    &  35.4~~~  &&  626 &  5.5 &&&  208 &  9.0   &&&  156 &  18.7 &\\
m    &  25.9~~~  &&  458 &  4.1 &&&  152 &  6.6   &&&  114 &  13.7 &\\
n    &  69.7~~~  && 1235 & 11.0 &&&  410 & 17.8   &&&  308 &  37.0 &\\
\hline
Total&&&&&&&&&&&&&\\
Mass:& 2.17~~~   && 37.2 & 0.256 &&& 17.8& 0.40   &&& 15.5 & 0.859 &\\
($10^9$M$_{\odot}$)&&&&&&&&&&&&&\\
\hline
\end{tabular}
\tablecomments{
The ({\it burst only}) column assumes that the entire luminous mass of each structure
component is from a single burst population, with age indicated by its
$F555W-F814W$ colors (see Table~\ref{arcages}). The following three
columns give the masses assuming an underlying redder population with
the spectrum of an Sb spiral galaxy superimposed with a fractional
burst population, with ages 0.001, 0.01 and 0.02 Gyrs
}
\end{table}

\clearpage

\begin{deluxetable}{llll}
\tablecaption{
Results of ellipse fits to the three prominent arcs around H1517+656. 
\label{ellipsefits}
}
\tablewidth{0pt}
\tablehead{
\colhead{Arc}&
\colhead{Radius (kpc)}&
\colhead{Ellipticity}&
\colhead{Sky Angle}}
\startdata
northwest&$15.4 \pm 1.0$&0.034 $^{+0.035}_{-0.034}$ &15$^{\circ}$ $^{+6}_{-15}$\\[0.25cm]
southeast&$20.8 \pm 0.7$&0.043 $^{+0.05}_{-0.043}$  &17$^{\circ}$ $^{+8}_{-17}$\\[0.25cm]
south    &$17.0 \pm 2.0$&0.083 $^{+0.25}_{-0.083}$  &24$^{\circ}$ $^{+14}_{-24}$\\
\enddata
\end{deluxetable}

\clearpage

\clearpage

\clearpage

\clearpage

\clearpage

\clearpage

\clearpage

\clearpage

\clearpage

\end{document}